\documentclass[a4paper]{article}

\usepackage{INTERSPEECH2022,comment,url,xcolor}

\title{NTT Multi-Speaker ASR System for the DASR Task of  CHiME-8 Challenge}
\name{Naoyuki Kamo$^*$, Naohiro Tawara$^*$, Atsushi Ando, Takatomo Kano, Hiroshi Sato, Rintaro Ikeshita, \\Takafumi Moriya, Shota Horiguchi, Kohei Matsuura, Atsunori Ogawa, \\Alexis Plaquet, Takanori Ashihara, Tsubasa Ochiai, Masato Mimura, Marc Delcroix, \\Tomohiro Nakatani, Taichi Asami, Shoko Araki
\thanks{$^*$Equal contribution}}

\address{NTT Corporation, Japan}
\email{\{naoyuki.kamo, naohiro.tawara, atsushi.ando, hrs.sato, marc.delcroix\}@ntt.com}
\begin{document}

\maketitle
\begin{abstract}
We present a distant automatic speech recognition (DASR) system developed for the CHiME-8 DASR track. It consists of a diarization first pipeline. For diarization, we use end-to-end diarization with vector clustering (EEND-VC) followed by target speaker voice activity detection (TS-VAD) refinement. To deal with various numbers of speakers, we developed a new multi-channel speaker counting approach. We then apply guided source separation (GSS) with several improvements to the baseline system. Finally, we perform ASR using a combination of systems built from strong pre-trained models. Our proposed system achieves a macro tcpWER of 21.3 \% on the dev set, which is a 57 \% relative improvement over the baseline.

\end{abstract}

\section{Introduction}
The distant automatic speech recognition (DASR) problem consists of identifying when each speaker speaks (diarization) and transcribing their speech (ASR) in conversations captured by distant microphones.
The CHiME challenge series has proposed tasks with increased levels of difficulty to measure progress in DASR, such as recordings of up to four speakers in home environments. 
The CHiME-8 DASR \cite{chime8-task1} track extends the difficulties of the previous editions by expanding the variety in the number of speakers per recording (up to eight), microphone array configurations, recording conditions, and speaking styles. 
Concretely, this is realized by requiring building a single DASR system, which can operate on four datasets, including dinner party recordings with four participants (CHiME 6 (CH6) \cite{chime6} and DiPCO (DiP) \cite{Segbroeck2019}), two-speaker interviews (Mixer 6 (MX6) \cite{brandschain2010mixer}) and a new corpus of business-like meetings called NOTSOFAR (NSF) \cite{notsofar-1}.

\begin{figure}[t]
  \centering
\includegraphics[width=0.70\linewidth]{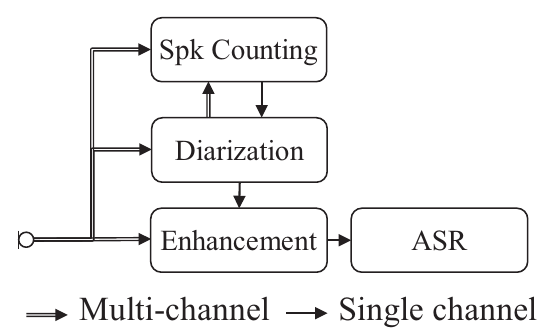}
  \caption{Proposed recognition system for DASR track.}
  \label{fig:pipeline}
\end{figure}

Our contribution to the CHiME-8 DASR track consists of a diarization first pipeline \cite{Raj2021}, which combines speaker diarization, speech enhancement (SE), and ASR as shown in Fig. \ref{fig:pipeline}. For the diarization, we extended our previously proposed end-to-end diarization with vector clustering (EEND-VC)-based diarization to include target-speaker voice activity detection (TS-VAD)-based refinement \cite{Medennikov2020TargetSpeakerVA,yang2024neural}. Besides, we developed a novel multi-microphone speaker counting approach, which estimates the number of speakers via speaker embedding clustering per microphone and combines the result across all microphones. The speaker counting is crucial for CHiME-8 DASR task as there is great variety in the number of speakers per recording, and wrongly estimating the number of speakers greatly impacts diarization and ASR performance.


For SE, we made several key modifications to the baseline guided source separation (GSS).
First, we propose a new rule for microphone subset selection, which is based on the envelope variance \cite{WOLF2014170} and the speech clarity index $C_{50}$ \cite{ISO3382-1}.
Besides, we refined the SE frontend by replacing the MVDR beamformer with the spatial-prediction multichannel Wiener filter (SP-MWF) \cite{benesty2008noncausal,cornelis2011performance}.
By doing so, we mainly aim to select a more effective reference microphone for beamforming, which is essential when dealing with distributed microphone arrays.

For ASR, we exploit the availability of strong pre-trained models, including Whisper, NeMo, and WavLM. First, we investigated fine-tuning Whisper and NeMo models on the CHiME-8 training data. We introduce a curriculum learning scheme to efficiently fine-tune Whisper on the very noisy CHiME-8 training data. 
In addition to the above models, we also developed a transducer-based ASR system, which uses WavLM as the front-end. This last model, although being much more computationally efficient, achieves comparable performance to the Whisper- and NeMo-based models.
Finally, we perform N-best rescoring and system combination.


In the remainder, we describe the different parts of our system, i.e., diarization and speaker counting in Section \ref{sec:diarization}, SE in Section \ref{sec:se} and ASR in Section \ref{sec:asr}. We then present overall results and analysis in Section \ref{sec:experiments}.



\section{Diarization and speaker counting}
\label{sec:diarization}
\begin{figure}[t]
  \centering
\includegraphics[width=1.0\linewidth]{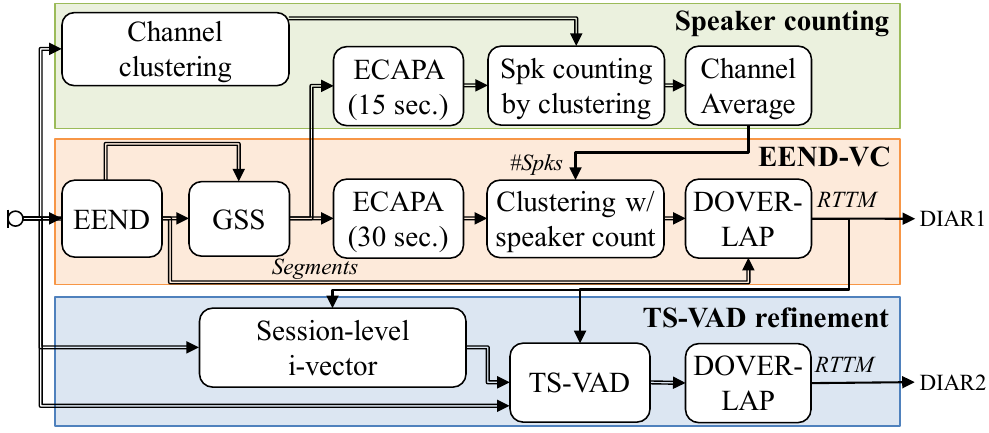}
  \caption{Proposed diarization and speaker counting system.}
  \label{fig:diarization}
\end{figure}
Figure \ref{fig:diarization} shows our proposed diarization system, which consists of EEND-VC \cite{EEND-vector-clustering_Interspeech2021} that relies on multi-channel speaker counting, followed with TS-VAD-based refinement \cite{yang2024neural}. 

\subsection{EEND-VC segmentation (DIA1)}
The EEND-VC module is based on our CHiME-7 submission \cite{kamo23_chime,tawara_icassp24}, with some essential modifications.
EEND-VC performs chunk-level segmentation to estimate the activity of each speaker in each chunk with EEND, where the maximum number of speakers in a chunk is set to $N^{\text{max}} = 4$. 
Next, we aggregate the chunk-level segmentation results by clustering speaker embeddings computed for each speaker in each chunk with the ECAPA-TDNN model\cite{desplanques2020ecapa}. 
This process is performed for each channel, and then combined with diarization output voting error reduction + Overlap (DOVER-Lap) \cite{Raj2021Doverlap}.

We made two main additions to our previous EEND-VC system. First, we employ GSS after the segmentation to obtain cleaner speaker embeddings for speaker counting and clustering. Here, we perform GSS using the segmentation obtained from each channel, which generates thus as many signals as channels. This allows us to generate multiple diarization outputs for DOVER-Lap, whose reliability may vary dynamically depending on the microphone used. 
Second, we perform constrained spectral clustering \cite{EEND-vector-clustering_Interspeech2021} by setting the number of clusters to the number of speakers obtained from the speaker counting module, described in the next subsection. 

\noindent \textbf{Settings:} 
We used the same configuration for EEND-VC as in our CHiME-7 submission \cite{tawara_icassp24,kamo23_chime}, which uses pre-trained WavLM-large \cite{WavLM} to obtain the input speech features. The model has approximately 324 million parameters. Different from our previous system, we use a chunk size of 30 seconds instead of 80 sec. We could obtain reliable embeddings with shorter chunks by using GSS to reduce the influence of the interference speakers. Reducing the chunk size is essential to allow dealing with recording with more than $N^{\text{max}}=4$ speakers with high overlap as observed in the NOTSOFAR data \cite{notsofar-1}.

\subsection{Multi-channel speaker counting}
It is challenging in short sessions to count speakers since the number of segment-level speaker embeddings is limited.
To solve this problem, we propose a speaker counting scheme that predicts the number of speakers from multi-channel signals.
Channel clustering and GSS are employed to mitigate the influence of spatial characteristic differences across channels on speaker embeddings.
The number of speakers is estimated in each channel cluster, enabling to increase the number of embeddings, which leads to better speaker counting performance.

First, input channels are clustered to find nearby microphone groups in the input session.
For this, we grouped channels using agglomerative hierarchical clustering based on inter-channel correlations from the initial fixed-length raw signals.
Then, we compute speaker embeddings on the output of GSS applied in the EEND-VC stage using the ECAPA-TDNN model.
For each microphone group, the number of speakers is estimated from all embeddings in the group using the approach proposed for normalized maximum eigengap spectral clustering (NMESC) \cite{park2020autotuning}.
Finally, the estimated speaker counts are integrated to obtain the session-level speaker counts $\bar{c}$, 
$\bar{c} = \lfloor \frac{1}{N} \sum_i n_i c_i \rceil$, where $c_i$ and $n_i$ are estimated speaker counts and the number of embeddings in the $i$-th microphone group, $N$ is the total number of embeddings in the session, and $\lfloor\cdot\rceil$ is a rounding function to an integer.
$n_i$ serves as a weight factor that emphasizes the estimations from groups with more microphones, which are thus more reliable.  

\noindent \textbf{Settings:} We divided the output of GSS into 15-second segments to generate more samples. 
We used the first 120 seconds of the signals and a correlation threshold of 0.3 for channel clustering. 
\subsection{TS-VAD refinement (DIA2)}
We apply memory-aware multi-speaker embedding with sequence-to-sequence architecture (NSD-MS2S) -based TS-VAD \cite{yang2024neural,wan23_chime} to refine the diarization results obtained with EEND-VC. NSD-MS2S exploits session-level ivectors, obtained by averaging the segment-level ivectors belonging to the same speaker. 
It then combines these ivectors with local speaker embeddings derived from the input mixture and the local segmentation information. 
These combined embeddings are used to condition a conformer-based TS-VAD module.
We use the same model configuration that was used in the CHiME-7 \cite{wan23_chime}, but with a stronger initial diarization provided by EEND-VC.


\noindent \textbf{Settings:} We rely on the publicly available implementation of NSD-MS2S for the TS-VAD refinement\footnote{\url{https://github.com/liyunlongaaa/NSD-MS2S}}. We used the default parameters, except that we used only a single deep interactive module block. The model has 5.80 million parameters. 

\begin{figure}[t]
  \centering
\includegraphics[width=0.99\linewidth]{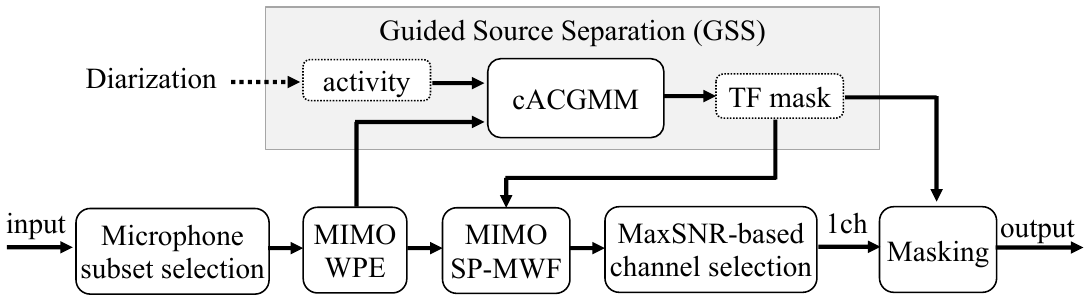}
  \caption{Our GSS-based SE frontend.}
  \label{fig:gss}
\end{figure}

\section{Speech enhancement (SE) front-end}
\label{sec:se}

Figure \ref{fig:gss} shows our proposed GSS-based SE frontend, which basically follows the official SE frontend \cite{boeddecker18_chime} implemented in the CHiME-8 DASR NeMo Baseline system \cite{chime8-task1,park23_chime}.
It consists of microphone subset selection, weighted prediction error (WPE)-based dereverberation \cite{yoshioka2012generalization,nakatani2010speech}, time-frequency (TF) mask estimation using cACGMM \cite{ito2016complex} based GSS \cite{boeddecker18_chime}, and SE using a variant of the multichannel Wiener filter (MWF) followed by TF masking.
Changes from the official SE frontend include improvement of microphone subset selection (Section \ref{sec:se:mic-selection})
and replacement of the MVDR beamformer (Section \ref{sec:se:bf}).

\subsection{Microphone subset selection}
\label{sec:se:mic-selection}

To select an effective subset of microphones for the SE frontend, we relied on two acoustic features:
the envelop variance (EV) \cite{WOLF2014170} (which is also used in the CHiME-8 DASR NeMo Baseline system)
and a speech clarity index $C_{50}$ \cite{ISO3382-1}.
The index $C_{50}$ is defined as the ratio of the energy in the early phase (0 to 50 ms) to that in the late phase (more than 50 ms) of the room impulse response.
We estimated $C_{50}$ using the Brouhaha toolkit \cite{lavechin2022brouhaha}.

We measured EV and $C_{50}$ for each microphone observation signal to rank the microphones.
Let $I_{\mathrm{EV}}$ (resp.\ $I_{C_{50}}$) be the set of the top $K$ microphones ranked by EV (resp.\ $C_{50}$),
where $K = 65$ [\%] in our setup.
Let also $I = I_{\mathrm{EV}} \cap I_{C_{50}}$ be the intersection of the two subsets.
We selected the subset of microphones to pass to the subsequent SE frontend as follows:
\begin{itemize}
  \item If $|I| \geq 15$, then we select $I$.
  \item If $|I| < 15$ and $I_{\mathrm{EV}} \geq 15$, then we select $I_{\mathrm{EV}}$.
  \item If $|I| < 15$ and $I_{\mathrm{EV}} < 15$, then we select the set of the top 15 microphones ranked by EV.
  \item We use all microphones when there are fewer than 15.
\end{itemize}
In the aforementioned rule, we assumed that using at least 15 microphones helps improve SE performance.


\subsection{Mask-based MIMO source separation filter}
\label{sec:se:bf}

As a source separation filter, the official SE frontend uses the mask-based MIMO MVDR beamformer with maximum SNR (MaxSNR)-based reference channel selection.
We replaced this beamformer part with the so-called spatial-prediction multichannel Wiener filter (SP-MWF) \cite{benesty2008noncausal,cornelis2011performance} given by
\begin{align*}
    \mathbf{w}_{f}(r) = 
    \frac{
        ( \mathbf{e}_r^\top \mathbf{R}_{\mathbf{x},f} \mathbf{e}_r )
        \mathbf{R}_{\mathbf{n},f}^{-1} \mathbf{R}_{\mathbf{x},f} \mathbf{e}_r
    }{
        \mu \mathbf{e}_r^\top \mathbf{R}_{\mathbf{x},f} \mathbf{e}_r
        + \operatorname{Tr}(
            \mathbf{R}_{\mathbf{n},f}^{-1} \mathbf{R}_{\mathbf{x},f}
            \mathbf{e}_r \mathbf{e}_r^\top
            \mathbf{R}_{\mathbf{x},f}
        ) 
    } \in \mathbb{C}^M,
\end{align*}
where $r \in \{ 1, \ldots, M \}$ denotes the reference channel,
$\mathbf{e}_r \in \mathbb{C}^M$ is the unit vector that selects the $r$-th microphone,
$\mu \in \mathbb{R}_{\geq 0}$ is a hypterparameter (we set it to $\mu = 0$),
and $\mathbf{R}_{\mathbf{x},f}$ and $\mathbf{R}_{\mathbf{n},f}$ are
the target-source and noise covariance matrices at frequency bin $f$.
The covariance matrices are estimated using the TF mask in the same way as in the baseline.
Finally, unlike the baseline, we do not apply blind analytic normalization (BAN) postfilter.


\section{ASR}
\label{sec:asr}
\begin{figure}[t]
  \centering
\includegraphics[width=0.9\linewidth]{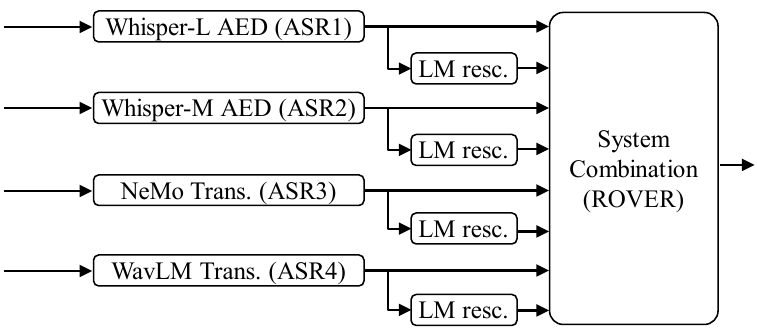}
  \caption{Proposed ASR backend.}
  \label{fig:asr}
\end{figure}

\subsection{ASR models}
\label{lab:ASR}
Figure~\ref{fig:asr} shows the schematic diagram of our ASR backend. We employed four ASR models: two attentional encoder-decoder (AED) models and two transducer models \cite{graves2012rnnt}.  Each ASR model generated N-best hypotheses, which were accumulated and rescored by language models. Both the beam size and N-best size were set to 4. Then, the best hypothesis was determined among the original and rescored N-best hypotheses using recognizer output voting error reduction (ROVER) \cite{Fiscus1997ROVER}. The architecture of our ASR models is described below.

\noindent \textbf{Whisper Large v3 (ASR1):}
We fine-tuned the Whisper Large V3 model \cite{radford2022whisper} for the CHiME-8 task. The model has 1540M parameters, i.e., 32 Transformer encoder-decoder layers with 8 attention heads and a hidden dimension of 1280. 
The vocabulary size was 51,864 with the GPT-2 \cite{radford2019language} byte-level BPE tokenizer.
The CHiME-8 training data is very noisy and thus unreliable for fine-tuning. To mitigate this issue, we proposed a curriculum learning scheme, which filters out utterances with a high character error rate (CER). Practically, during fine-tuning, we changed the target transcription of the cross-entropy loss to the self-generated decoding result when the CER exceeds 30~\%. In that case, we multiplied the loss by 1/1000th to reduce its impact. Otherwise, we used the ground-truth transcription as a reference. By computing the CER on the fly, we can adapt the number of training data as fine-tuning progresses, i.e., as the model becomes stronger, the CER decreases, and more difficult data can be reliably used.

\noindent \textbf{Whisper Medium (ASR2):}
We also utilized Whisper Medium English model \cite{radford2022whisper} to initialize a Transformer-based encoder-decoder model. 
This model had approximately 770M parameters and consisted of a 24-layer Transformer encoder and decoder, each with 8 attention heads and a model width of 1024. 
The vocabulary size was 51,864 with the GPT-2 \cite{radford2019language} byte-level BPE tokenizer.
Despite having significantly fewer parameters than Whisper Large V3, it is pre-trained solely on English data, making it potentially more suitable for the CHiME-8 tasks. 

\noindent \textbf{NeMo Transducer (ASR3):} 
We adopted an official pre-trained NeMo transducer model, which had 644M parameters, and finetuned it using the CHiME-8 dataset. The NeMo transducer model consists of two-layer 2D convolutional neural networks (CNNs) followed by 24 fast conformer blocks \cite{anmol2020conformer,rekesh2023fastconformer}. The prediction and joint networks had a 640-dimensional long short-term memory (LSTM) and a 640-dimensional feed-forward network. The number of output units was 1025 byte pair encoding (BPE) tokens. 

\noindent \textbf{WavLM Transducer (ASR4):} 
We built another transducer-based ASR system that uses the weighted sum of WavLM \cite{WavLM} Transformer layers as input features. 
The ASR encoder has two-layer 2D-CNNs followed by 18 branchformer blocks \cite{kim2022branchformer}. The prediction and joint networks had two-layer 640-dimensional LSTMs and a 512-dimensional feed-forward network, respectively. We adopted 500 BPE tokens as output units. The total number of parameters was approximately 422M.
We conducted three-step training to build the ASR4 system sequentially: 1) partial parameters were trained using the CHiME\&LibriSpeech\&VoxCeleb datasets while freezing WavLM front-end, 2) all network parameters (including WavLM) were fine-tuned using the same data from the first step, and 3) we fine-tuned it using only the CHiME-8 data. 

\noindent \textbf{LM:} We built a 35M parameters of Transformer-LM for LM rescoring. The LM has the vocabulary of 1000 BPE tokens. We pre-trained the LM using 1/10 of the LibriSpeech text dataset and then fine-tuned using the CHiME-8 train text dataset. At the inference, the LM uses 256 past rescored (re-ranked) 1-best tokens as the context (i.e., context carry-over) \cite{Ogawa_arXiv2024}. 

\section{Training data}
\noindent \textbf{Diarization:}
Each diarization model (EEND-VC and TS-VAD) was initially trained using simulated mixtures and then finetuned using the CHiME-8 training set.
The protocol for simulating mixtures basically followed the method that attempts to make utterance transitions natural \cite{yamashita2022improving}, but the following modifications were made to have more similar statistics to the real data: i) we first considered turn-hold, turn-switch, and interruption to generate long-form audio, and then inserted backchannels afterward, ii) we directly sampled durations of silence/overlap between utterance from the real data instead of sampling from any fitted distribution, and iii) overlap durations in interruptions/backchannels were determined from absolute durations extracted from the real data (instead of relative ratios).
We generated 1M and 91k 50-second mixtures of four speakers using LibriSpeech \cite{panayotov2015librispeech} for training EEND-VC and TS-VAD, respectively.
We used 500 mixtures for validation in both cases.
Each mixture was augmented using the simulated room impulse responses \cite{ko2017study} and MUSAN noises \cite{snyder2015musan}.


\noindent \textbf{ASR:}
We used 70 hours of CHiME-8 training data processed with GSS for the Oracle segmentation. 
We did not use train\_call and train\_intv in Mixer6 at this time because preliminary experiments showed minimal improvements with them.
For the pre-training of the transducer-based ASR model with WavLM (ASR4), we also used LibriSpeech \cite{panayotov2015librispeech} and VoxCeleb1+2 \cite{voxceleb} datasets in addition to the CHiME-8 described above. Note that the contrastive data selection algorithm \cite{lu22_interspeech,kamo23_chime} was applied in an unsupervised fashion to the unlabeled VoxCeleb1+2 data, reducing its size by a quarter. We utilized the dev set for early stopping with a patience of 5 epochs.

\section{Experiments}
\label{sec:experiments}
We report brief analysis of the different components of our system followed by the overall results in Section \ref{sec:overal_res}
\subsection{Analysis I: Diarization and speaker counting}
Table \ref{tab:diarization_results} shows the diarization error rate (DER) on dev set, without and with TS-VAD refinement, DIA1 and 2, respectively. Both systems greatly outperform the baseline.
Note that we used a similar TS-VAD model as that of the top CHiME-7 system \cite{wan23_chime}, but with a more powerful initialization with EEND-VC. Consequently, we achieved better performance with a single TS-VAD refinement pass, i.e., DERs of 24.0~\% and 6.1~\% on CHiME-6 and Mixer-6 dev set, compared to 25.8~\% and 8.9~\% with 4 diarization pass in the CHiME-7 top system \cite{wan23_chime}.

Table \ref{tab:counting_results} shows the channel-wise and microphone group-wise speaker counting accuracy, without and with group averaging. Our proposed system accurately estimated the number of speakers for CHiME-6, DiPCO, and Mixer 6 datasets. For NOTSOFAR, we achieved a lower accuracy of 58.2 \%  accuracy because there is more variability in the number of speakers, and the recordings are short. The proposed speaker counting approach greatly outperforms the baseline in all conditions.

\begin{table}[t]
  \caption{DER [\%] $ (\downarrow$) on dev set computed with md-eval with a collar of 0.25 sec.}
  \vspace{-3mm}
  \label{tab:diarization_results}
  \centering
  \resizebox{\linewidth}{!}{%
  \begin{tabular}{@{}l@{\hspace{0.2cm}}lc@{\hspace{0.2cm}}c@{\hspace{0.2cm}}c@{\hspace{0.2cm}}c@{\hspace{0.2cm}}c@{}}
    \toprule
    ID & Model & CH6 &DiP & MX6 &NSF & Macro\\
    \midrule
DIA0 & Baseline (NeMo)&  45.65 	&	45.92 	&	25.16 & 38.05 & 38.70 \\
\midrule
DIA1 & EEND-VC w/ ECAPA & 28.52 	&	24.38 	&	9.69 	&	10.67 	&	18.32 \\ 
DIA2 & + TS-VAD & 23.97 	&	21.01 	&	6.11 	&	9.72 	&	15.20  \\ 
      \bottomrule
  \end{tabular}}%
  \vspace{-3mm}
\end{table}

\begin{table}[t]
  \caption{Speaker counting accuracy [\%] ($\uparrow $) on the dev set.}
  \vspace{-3mm}
  \label{tab:counting_results}
  \centering
  \resizebox{\linewidth}{!}{%
  \begin{tabular}{@{}l@{\hspace{0.2cm}}c@{\hspace{0.2cm}}c@{\hspace{0.2cm}}c@{\hspace{0.2cm}}c@{\hspace{0.2cm}}c}
    \toprule
    & CH6 &DiP & MX6 &NSF & Macro\\
    \midrule
Baseline (NeMo) & 50.0 & 0.0 & 100.0 & 13.8 & 41.0 \\
\midrule
Channel-wise counting & 95.5 & 84.3 & 99.7 & 48.5 & 82.0 \\
Microphone group-wise counting & 100.0 & 90.0 & 100.0 & 57.5 & 86.9 \\
~+  Group averaging & 100.0 & 100.0 & 100.0 & 58.2 & 89.6 \\
      \bottomrule
  \end{tabular}}%
\end{table}

\subsection{Analysis II: Speech enhancement}

In preliminary experiments, 
we confirmed that our new microphone subset selection (Section \ref{sec:se:mic-selection}) could improve the tcpWER from 20.00\% to 19.41\% in the CHiME-6 dataset and from 31.43\% to 29.70\% in DiPCo when using the baseline NeMo ASR system with oracle diarization.
Note that since Mixer 6 and NOTSOFAR have fewer than 15 microphones, microphone subset selection was not applied to these two datasets.
We also observed that replacing the separation filter and turning off the postfilter in the baseline system could improve the macro tcpWER by more than 0.3\%.

\subsection{Analysis III: ASR}

\begin{table}[tb]
  \caption{tcpWER [\%] ($\downarrow$) on the dev set with oracle diarization and SE front-end.}
  \vspace{-3mm}
  \label{tab:asr_ablation}
  \centering
  \resizebox{1.00\linewidth}{!}{%
  \begin{tabular}{@{}l@{\hspace{0.2cm}}lc@{\hspace{0.2cm}}c@{\hspace{0.2cm}}c@{\hspace{0.2cm}}cc@{}}
    \toprule
     
    ID & Model & CH6 & DiP & MX6 & NSF & Macro \\
    \midrule
    ASR0 & NeMo Trans. (Baseline) &  19.78 & 31.01 & 10.61 & 17.95 & 19.84 \\ 
    \midrule
    ASR1 & Whisper-L AED & 17.80 & 26.29 &	10.43 & 13.05 & 16.89 \\ 
    ASR2 & Whisper-M AED & 19.81 & 27.15 & 11.16 & 13.57 & 17.92 \\ 
    ASR3 & NeMo Trans.    & 20.30 & 28.33 & 11.25 & 14.33 & 18.55 \\ 
    ASR4 & WavLM Trans.    & 19.76 &	27.52 &	10.79 &	13.23 & 17.82 \\ 
    \midrule
   ASR5 & ROVER (ASR $\times$ 6 +LM resc.) & 16.42	& 23.71	&9.42	&	11.44	& 15.25 \\
    \bottomrule
  \end{tabular}%
  }
\end{table}

Table \ref{tab:asr_ablation} shows the tcpWERs of our developed ASR backends when using Oracle diarization and our SE front-end. 
The tcpWERs of all systems were significantly improved compared to those of the baseline. Although the best single ASR system is the Whisper Large v3 (ASR1), the WavLM transducer (ASR4), which is the smallest system, achieved comparable performance. 
Note that we trained two versions of ASR1 and ASR4 with different training hyper-parameters, but only reported the best version in Table \ref{tab:asr_ablation}. 
ASR 5 consists of the combination of the six ASR systems using both 1-best and hypothesis obtained after LM rescoring.

\begin{table}[tb]
  \caption{tcpWER [\%] ($\downarrow$) on the dev set. The real-time factor (RTF) is computed on the NOTSOFAR dev set.}
  \vspace{-3mm}
  \label{tab:overall_res}
  \centering
  \resizebox{\linewidth}{!}{%
  \begin{tabular}{@{}l@{\hspace{0.2cm}}l@{\hspace{0.2cm}}l@{\hspace{0.2cm}}l@{\hspace{0.2cm}}cccccc@{}}
    \toprule
    ID & Diar & SE & ASR & CH6 & DiP& MX6 & NSF & Macro & RTF\\
  \midrule
      \multicolumn{4}{c}{Baseline (NeMo)}  &  49.29 &	78.87 &	15.75 &	56.21 & 50.03  & - \\
    \midrule
    NTT-1 & DIA1  & SE  & ASR4  & 30.14	&	35.86 & 10.94	&	23.85 &25.20	& 2.46 \\ 
    NTT-2 & DIA2  & SE  & ASR1          & 28.21 &	35.32	& 10.66	&	20.41	& 23.65 & 3.14 \\ 
     NTT-3 & DIA2  & SE  & ASR5 (ROVER)          &25.49 &	31.25 &	9.63 &		18.79 &	21.29 & 4.03\\
    \bottomrule
  \end{tabular}%
  }
\end{table}

\subsection{Overall results on dev set}
\label{sec:overal_res}
 Table \ref{tab:overall_res} compares the results of the three proposed DASR systems (NTT-1--3) with the baseline.
The systems are sorted by increasing order of complexity.

\noindent \textbf{NTT-1} is our most lightweight model, which uses only EEND-VC for diarization (without TS-VAD refinement), our proposed GSS front-end with microphone selection (SE1), and our most efficient ASR system (ASR4).

\noindent \textbf{NTT-2} uses a stronger diarization (DIAR2) and ASR system (ASR1). 

\noindent \textbf{NTT-3} performs decoding with six ASR backends and system combination with ROVER as ASR5 in Table \ref{tab:asr_ablation}.

Our full system (NTT-3) achieves a relative improvement of 57~\% over the baseline. Note that it also achieves a 40 \% relative tcpWER improvement over the DASR baseline, despite the fact that our system generalizes to more diverse recording conditions.
NTT-3 is relatively complex and involves the combination of several ASR backends. However, lighter versions (NTT-1 and NTT-2) of our system can still achieve more than 50~\% relative improvement compared to the DASR baseline. 

We also provide RTF values computed on one A6000 GPU on the NOTSOFAR dataset, but these numbers are only indicative as our code has not been optimized for computational efficiency.


\clearpage
\bibliographystyle{IEEEtran}
\bibliography{refs}


\end{document}